# Rethinking Network Connectivity in Rural Communities in Cameroon

Jean Louis FENDJI KEDIENG EBONGUE
University of Ngaoundéré, Ngaoundéré, Cameroon
Tel: +237677038781, Email: jlfendji@uni-ndere.cm

**Abstract:** To bridge the digital divide between the urban and rural regions, the government of Cameroon has launched the Multipurpose Community Telecentres (MCT). But this project does not seems to sustain the local development. The aim of this study is threefold: to determine the ICT penetration in rural Cameroon and Internet adoption; to evaluate the impact of MCTs in rural region and to provide some recommendations for both the network planning and the development of suitable services. The study considers two rural communities in Cameroon. The results show that despite low incomes, and MCTs that are missing their goal, there is some readiness of local populations to welcome ICT projects in order to improve their daily life and activities. To design sustainable ICT projects for those regions, we provide some recommendations from network design to business strategy.

## 1. Introduction

Cameroon is a country in central Africa that covers an area of 475 442 km² with a population estimated at 23 130 708 of inhabitants in 2014 [1]. Presently, several projects in the field of Posts, Telecommunication and Information and Communication Technologies are launching, among which: Undersea cables, Central African Backbone, Urban optical loops, NBN Project (National Broadband Network), E-Post Project, Pan-African network of online services and Multipurpose Community Telecentres [2]. Despite all these projects and a good percentage of Mobile telephone subscription, Cameroon is still experiencing a very low percentage of individuals using Internet. Table 1 shows that this percentage is around a third of the one of Africa [3].

*Table 1: Comparison of Telecommunication Indicators in Cameroon, Africa, and World*

|  | Cameroon | Africa | World |
|---|---|---|---|
| Fixed-telephone subscriptions | 3.59% | 1.30% | 16.20% |
| Mobile telephone subscriptions | 70.39% | 65.90% | 93.10% |
| Fixed (wired)-broadband subscriptions | 0.08% | 0.30% | 9.40% |
| Percentage of Individuals using Internet | 6.40% | 16.80% | 37.9% |

The distribution of this low penetration rate of Internet is mainly among urban regions; leaving rural regions underserved.

In 1998, competition in mobile telephony was introduced with the aim to provide network coverage throughout the country. But mobile operators, that are profit-driven, are looking for more attractive regions where the return on investment can be guaranteed. Therefore, the coverage of these rural regions is left to the responsibility of the government. In this context, the government has launched the Multipurpose Community Telecentres (MCTs) project, designed to be deployed in rural and sub-urban areas in order to reduce the digital divide between these areas generally neglected by private operators and urban areas.

A MCT is a public infrastructure for telecommunication services, computers and the Internet to enable access to telecommunication services and the Internet at a low cost to the whole local community. It should also be a support to all activities and projects of the area. [4]. The initial goal when launching this project in 2003 was to create 2000 MCTs by 2015 but to date (December 2014), only 170 MCTs are deployed throughout the country. Most of them are connected through VSAT technology, offering limited capacity with reduced service quality and extremely high costs. Initially, MCTs were intended to provide a wide-range of telecommunication services as illustrated in Figure 1. But in reality most of MCTs are offering only some services and are not really different from Internet café. The distribution of MCTs throughout the country in December 2014 is presented in Table 2.

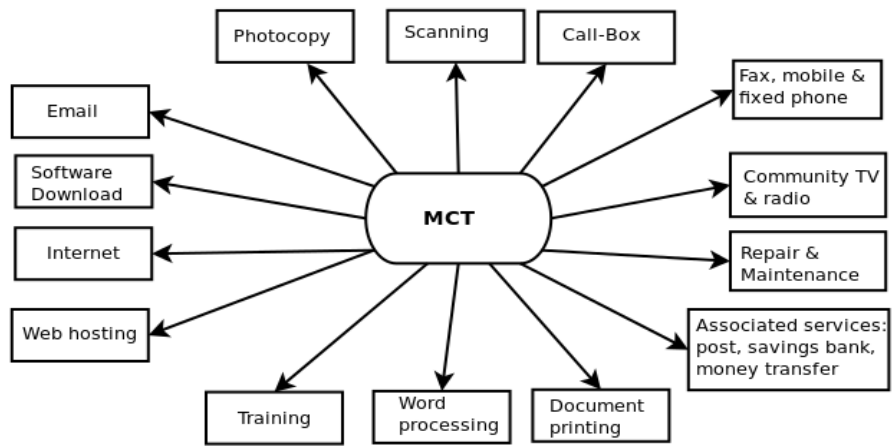

*Figure 1: MCT Services*

*Table 2: Distribution of MCTs in Cameroon*

| Region | **Adamawa** | **Centre** | **East** | **Far-North** | **Littoral** |
|---|---|---|---|---|---|
| Number of MCTs | 9 | 14 | 23 | 21 | 15 |
| Region | **North** | **North-West** | **South** | **South-West** | **West** |
| Number of MCTs | 17 | 17 | 24 | 7 | 22 |

Since the result of this project is very far from the expected outcomes, it is imperative to rethinking the rural connectivity. To achieve this, we need to evaluate the present situation and provides some recommendations.

## 2. Objectives

The main objective of this paper is threefold: to determine the ICT penetration in rural Cameroon and Internet adoption; to evaluate the impact of Multipurpose Community Telecentres (MCT) in rural region and to provide some recommendations for both the network planning and the development of suitable services.

## 3. Methodology

For this study, we selected two rural communities in Cameroon. The first is Mbé, located in the Adamawa region. The second is Garoua-Boulai which is located in the East. Because of lack of Internet infrastructure and the lack of skill using online survey tools, a questionnaire and personal face-to-face interview method was used. We designed a questionnaire for general public and two structured interviews. The survey is cross-sectional, meaning that the questionnaire is administrated to each sample member only once. The interviewer can affect the sample member's response. So to reduce this effect, the sample member completes the questionnaire themselves. A team member assists only if the sample member cannot read or write. At the end, the team member gets back the questionnaire.

The questionnaire is mainly composed of closed questions with multiple choices, in order to reduce the ambiguity and to facilitate the data analysis. Questions are grouped into four parts: questions 1 to 6 intend to describe the socio-economic situation; questions 7 to 10 aim to depict the ICT utilisation; questions 11 to 15 intend to portray the Telecentre impact and questions 16 to 18 aim to identify important E-services and the affordability of local population. The questionnaire is translated into two languages (English and French).

Two structured interview have been also defined: the first for those doing business and the second for telecentre managers. The aim of these interviews is to retrieve relevant information to help people doing business to improve their activities and also to learn from the experience of telecentre managers what can be done and what can be avoided.

## 4. Results from Data Analysis

202 persons responded to the questionnaire: 90 in the community of Mbé and 112 in the community of Garoua-Boulai.

*4.1 Socio-Economic Aspect*

Buyer/seller and civil servant are the main occupations in the studied areas with 35.14% of the sample. But we should note that most of the buyers/sellers are also farmers. They cultivate the bulk of the products that they are selling. A great part of civil servants, lecturers and workers in other sectors are also involved in agricultural activities during their spare time in order to supplement their monthly incomes. According to the Community Development Plan of Garoua-Boulaï [5] and that one of Mbé [6], 80% of the local population is involved in the agricultural sector.

Regarding the educational level, almost half of the sample declared to have a secondary level. Less than 10% did not go to school and around 13% have a primary level of education. Around 30% have at least the certificate for university entrance. This last category mainly consists of civil servants and lecturers.

Concerning the incomes, 20% do not mention their incomes. However, a third of the sample has no more than 25000 XAF (38 EUR) per month. That means less than 1.25 EUR per day. So the purchasing power is very low as it is the case in many other rural regions.

*4.2 Telecommunication Device Availability*

Half of the sample indicates to have access to a computer at home, at school, in the office, or somewhere else. Figure 2 shows that more than 92% have a telephone or a smartphone and around 8% own a tablet. Only less than 5% do not possess a telecommunication device. These figures indicate the fact that ICT is gradually penetrating the rural and suburban region. In another survey in rural South Africa, authors noted a penetration percentage of 98% of telephone [7]. This shows that developing applications and services targeting telephones may be the best option.

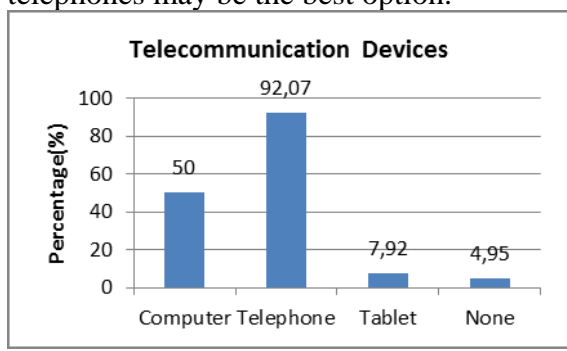
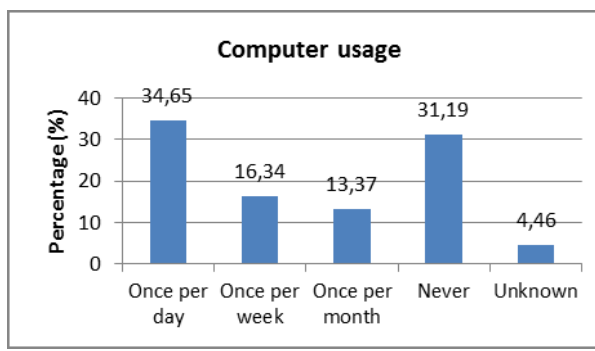

*Figure 2: Device availability*     *Figure 3: Computer usage*

Almost all computers run with Windows as operating system. The main reason is that these computers have been bought with a pre-installed Windows. The results also show that around 70% of telephones have a browser. However, numerous browsers offer just basic features using the Wireless Application Protocol (WAP) instead of real HTTP.

Regarding the computer usage, Figure 3 shows that a third of the sample has never used a computer while approximately the same percentage is using a computer on a daily base, and half at least once per week.

### 4.3   Internet Adoption

As it is shown in Figure 4, around half of the respondents indicated not having a personal opportunity to access the Internet. The majority of those who have access to Internet use only their telephones or smartphones. One the most important reasons are the costs of the equipment (USB dongle or router); the connexion fees are also very expensive for this social category. For example, offers from MTN [MTN 2014] are 350 XAF (0,53 EUR) per day with a limit of 150MBof data; 8000XAF (12,19 EUR) per month with a limit of 1GB. These offers slightly differ from those of Orange.

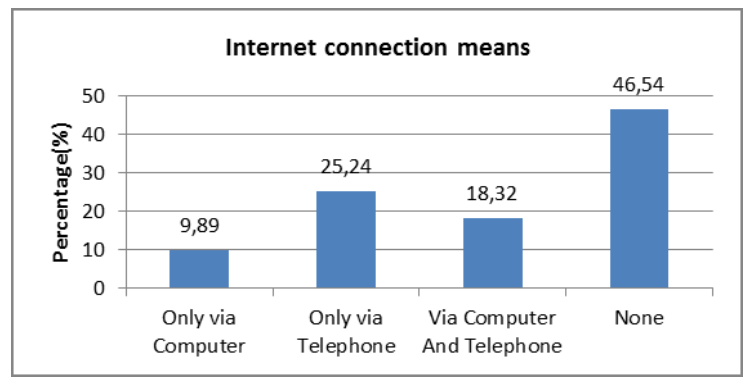

*Figure 4: Type of Internet Connection*

The Internet has been never used by around 40% of the sample because of the still very expensive costs. Nevertheless, just less than a quarter recognises using the Internet on a daily or weekly base.

### 4.4   Web and Social Network

As presented in Figure 5, the most used website is Google with 44% of the sample. Since it is a search engine, it is known by almost all of those who access the Internet. Yahoo is also known but usually as no more than a messaging service. With around 40%, Facebook is the most famous social network. With these figures and adding the fact that more than two third of the sample have telephones with a browser, developing web-based solutions for these regions may be more efficient.

Services providing voice over IP like Skype are not really known. Youtube is known by less than 10% of those accessing the Internet.

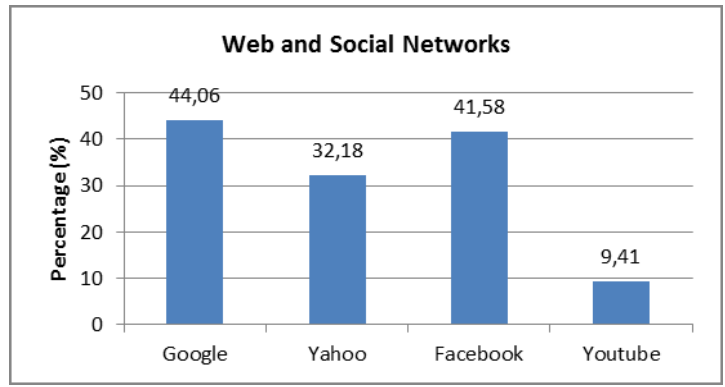

*Figure 5: Use of Web and Social Network*

### 4.5 Affordability

With very low incomes, around 40% of the sample are ready to pay no more than 1000XAF (1,52 EUR) per month. Only 26% declare to be able to pay at least 5000XAF (7,62 EUR) per month. This result shows also that only around 10% are able to afford the monthly offer of mobile operators (12,19 EUR). A monthly fee should be set to not more than 2500 XAF (3,80EUR) in order to allow the majority to afford the service. The affordability of the sample is presented in Figure 6.

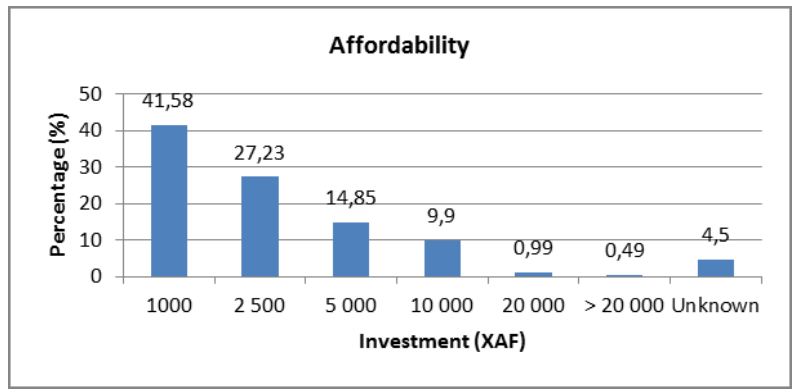

*Figure 6: Affordability*

### 4.6 MCT assessment

Figure 7 shows that MCT are located in the centre of the community close to the population. 40% mentioned that they are living in a radius of less than one kilometre around the MCT and a total percentage of 77% live not more than three kilometres away from it.

Despite the fact that more than two third of the sample are located at not more than three km from a MCT, 70% of the sample are never gone to it, as shown in Figure 8. A great number of clients spends just one hour in the MCT. They could spend more time but the cost of a working session is very expensive for them, around 300XAF (0,46EUR).

Figure 9 shows that the few number of MCT clients principally use it for education by performing some research in Google or accessing some online encyclopaedias like Wikipedia. Beside education, some people are using the MCT for working purposes. Keeping in touch with relatives is the concern of less than a quarter of the MCT clients.

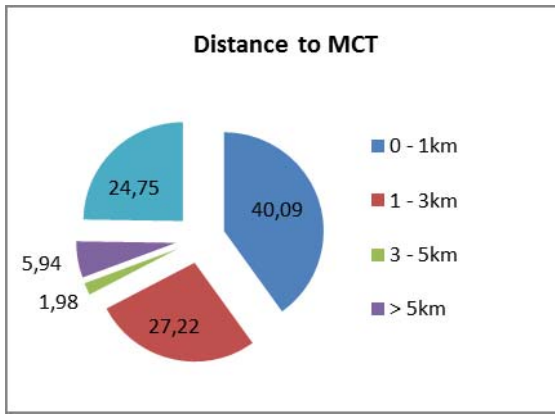
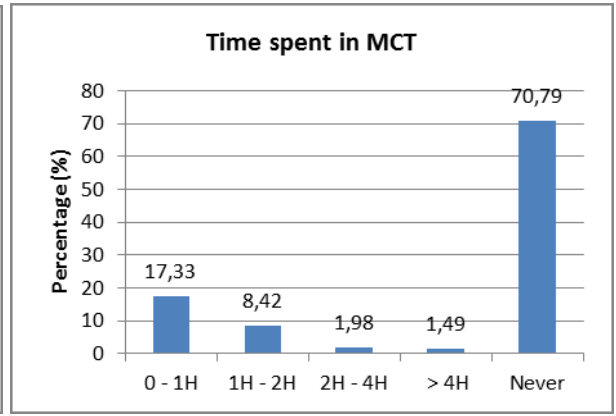

Figure 7: Distance to the MCT                     Figure 8: Time spent in MCT

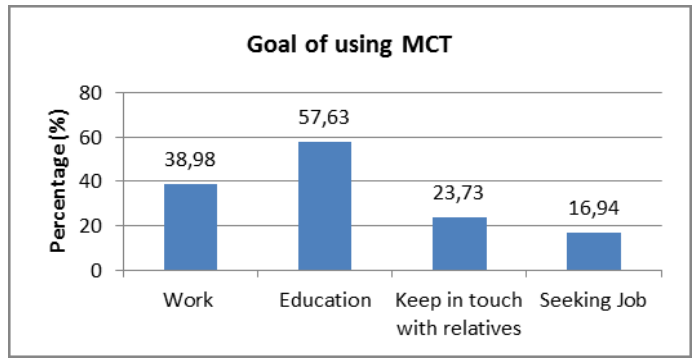

Figure 9: Goal of using MCT

### 4.7 Desired services and applications

As shown in Figure 10, e-education appears to be the most desired service. This is confirmed by the fact that education is the most important reason of using MCT. Moreover, the result is closed to the result from the ITU analysis on global survey on rural communication [8]. E-health is the second most important and e-administration the third one.

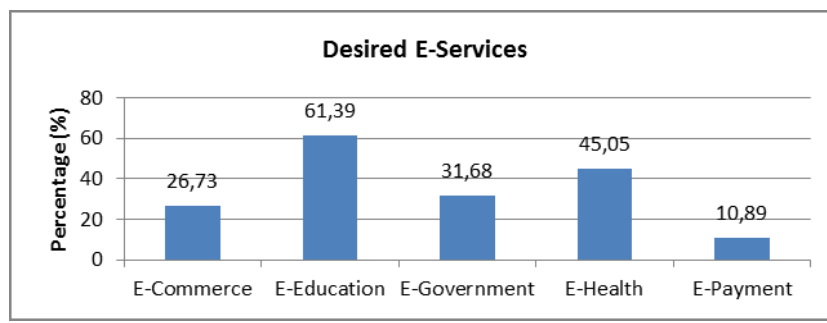

Figure 10: Desired E-Services

## 5. Recommendations

After analysing data from questionnaires and results from interviews, some recommendations could be pointed out and they could be classified into three main categories: Network design, Software utilisation and services development and Business strategy.

## 5.1 Network design

### 5.1.1 Covering Areas of Interest

Since rural areas in Cameroon as in many other African countries are generally characterised by scattered settlements, there is no need to cover a whole region. Only areas of interest should be covered to minimise deployment cost. Therefore, wireless mesh networks are more cost-efficient than the infrastructure of common national network operators.

An area of interest should satisfy some criteria:

- There is a need to be met or a problem to be solved by the network;
- Recipients should be able to access the network. That means they have the required compatible equipment;
- Recipients should be able to contribute to the sustainability of the network.

### 5.1.2 Lowering CAPEX: Using Off-the-Shelf Materials

To lowering capital expenditures, low cost materials should be used. For example routers and servers can be based on standard PC hardware using free and open source software [9].

### 5.1.3 Catching and Caching

Despite the fact that some MCTs are switched onto the optical fibre, the majority of MCTs will stay connected by VSAT technology, which provides a very small and costly bandwidth. There is therefore a crucial need to reduce external traffic, since the bandwidth is very small (256 or 512Kbps in downlink and half in up). Thus, the first attempt to optimise the bandwidth is to use a caching proxy. It reduces the bandwidth and improves response times of frequently requested web pages by caching and reusing them.

### 5.1.4 Distance Monitoring

Due to the lack of local capacity in network administration, it is important to allow distance monitoring [9]. This can be done by providing remote login to all network components by configuring free monitoring tools such as Icinga (backward compatible to Nagios), nmap, ntop.

### 5.1.5 Independent Powering

To overcome the challenge of powering devices, an independent source of powering should be used. Solar panels are very good candidates since the climate in many regions in Cameroon is most of the time sunny.

## 5.2 Applications and Services

### 5.2.1 Using Free Software

With 98% of computers using Windows as operating system, system update and updates of antivirus programs and other Windows applications could require huge amount of outgoing traffic as reported in [10].

Therefore, to save bandwidth and also to avoid illegal use of licensed applications, it is important to switch on free software. Free operating systems like Linux allow the creation of local repositories containing most of the packages we could need. By creating a local repository, programs are downloaded from the Internet to local repository once and from this they can be installed on any computer connected to the local network. Doing this, the bandwidth is saved, illegal use of licensed applications is avoided and problems due to malware and viruses especially important in Windows system are limited.

Switching to free software requires the sensitization for the benefits of using free software and the organisation of training sessions for local population.

### 5.2.2 Making Content Locally Available

Using local repositories and caching proxies is a good attempt to make a better use of bandwidth. Nevertheless, this can still be optimised by hosting useful online materials locally. So, the external traffic will be limited, contributing in saving bandwidth.

### 5.2.3 Coupling Web and SMS

Web content can be accessible via Wi-Fi and readable by a browser using smartphones or computers. Despite the fact that around 70% of telephones have a browser, these browsers offer just basic features by using WAP instead of real HTTP. Moreover almost all these telephones do not have a wireless interface such as Wi-Fi to directly communicate with the network. So to take advantage of WAP, we should use mobile data service such as General packet radio service (GPRS) and develop application using WML (Wireless Markup Language). But the lack of good authoring tools and the under-specification of terminals requirements do not foster the development of WAP applications. Therefore we should consider SMS and couple it to the Web in order to provide relevant information.

### 5.2.4 Elementary Services

We should think about elementary services rather than a big bundle when developing services. This is for two reasons:
- Not all services are relevant for all communities because main activities and interests are differing from a community to another one. Therefore, they should be developed separately. However, we should keep in mind that services could communicate.
- Since services should also be SMS-based (in addition to web service), it is important to develop elementary services.

## 5.3 Business Strategy

### 5.3.1 Creation of Local Market

The first reason why people do not access the Internet is not the costs but rather the lack of knowledge about the benefits of using the Internet [11]. Thus, the creation of a local market goes via information and sensitising of the population about the benefits of ICT and the Internet in their daily live. Presenting some success stories might help to stimulate some sceptics.

However, having the will is not enough. People should also have the capacity in using electronic services. Therefore, building capacity is of great importance.

### 5.3.2 Public Private People Partnership

The great beneficiary of building first mile solutions in rural areas is the government. In fact, most of the services and infrastructure deployed in rural areas are public because enterprises, which are looking for profit, are not interested in these regions. But the government alone cannot provide a sustainable first mile solution. The analysis of MCTs showed their limitation and the poor impact on the local population. The government should be accompanied in this task by the private sector and the local population. For synergy effects it is important to point out the role, which each stakeholder might play.

- Public: the government should provide the basic infrastructure, as it is the case with MCT. Moreover it should reconsider the legal framework to allow the use of some licensed frequencies in those areas. For example, some studies have shown the capacity of low cost solutions like OpenBTS to provide communication via GSM frequency [12], but this type of initiative always faces the problem of licensed bands.
- Private: Private enterprises could be involved in the development of services or the extension of the network. For example supporting third party SMS-based services.
- People: they could play an important role in the generation and sharing of local content and information.

### 5.3.3 Different Access: Local and Internet

The costs of accessing the Internet are still very high even for those living in urban and suburban Cameroon. In [13] authors argued that the sustainability of this type of project depends on the pricing structure. With the low capacity and the high costs of the Internet in rural regions, it is important to provide different access possibilities:
- Bronze Package -Local access: the user can access local services and local content at very low costs;
- Silver Package - Local access + specific online services: the user can access local services and local content + some specific online applications, for example to retrieve market prices of product;
- Gold Package - Local access + Internet: the user has access to local services and local content + Internet.

## 6. Expected Benefits

The emerging policy of Cameroon relies largely on activities performed in rural regions like agriculture, breeding, fisheries. For example, with a contribution of 20.6% in GDP in 2013 not too far from industry 27.3% [14], the agricultural sector is a real pillar of the national development. Redefining the rural connectivity in this way would be very helpful to improve local activities by helping inhabitant accessing useful information. For example, farmers could have information about market prices or weather; they could also look for funding and obtain seed and fertilizers at very low prices. In the same direction, educational sector can really be improved. Local schools are experiencing very high failure rates in official exams. The pass rate to the Primary School Leaving Certificate was 11% in Mbé in 2012 [6]. The main reason of this failure is the lack of qualified teachers. As in [10], making offline contents, such as offline Wikipedia, locally available could help to improve educational sector. Additionally, the health sector could to be supported. As everywhere else in rural regions, this sector is facing the absence of doctors, shortage of skilled staff and infrastructure and difficulties in drugs supply. Accessing information could help to sensitize local population about diseases and provide some recommendations for auto-medication or in case of emergency. In summary, new opportunities could be provided both for populations, local institutions and organisms by transforming MCTs as rural Hub.

## 7. Conclusions

In this paper we presented MCT project designed by the Cameroonian government with the aim of reducing the digital gap between rural and urban Cameroon, and we also portrayed the ICT penetration in two rural communities. The results showed that the impact of MCTs on local populations is very low; despite the readiness of local populations to adopt ICT projects in their daily life, and the potential of these projects to boost the local development. As mentioned in [15], it is cleared that network design in rural regions is not just a technical matter. MCTs should be turned into rural wireless community in order to support relevant sectors like agriculture, education and health and therefore enhance the local development. To provide this first mile solution with a particular consideration of sustainability, we provided some recommendations in three plans: network design, software utilisation and services development, and Business model.

A low cost solution for covering only areas of interest while lowering CAPEX is wireless mesh network. Although a design methodology for this type of network in rural regions is provided in [15], determining the optimal positions of the nodes of the network in respect to technical, environmental and socio-economic constraints is still a challenging task. Another interesting future work is to find a way to couple the Web and SMS. How to develop application targeting both web and SMS users? Providing such a framework will allow simple telephone to retrieve via SMS relevant information from advanced web services.